\documentclass[9pt]{article}
\usepackage[preprint]{spconf}
\usepackage[latin9]{inputenc}
\usepackage{amsmath}
\usepackage{amssymb}
\usepackage{graphicx}
\usepackage{cite}

\copyrightnotice{978-1-5386-8151-0/18/\$31.00~\copyright2018 IEEE}

\name{Ryan M. Corey, Naoki Tsuda, and Andrew C. Singer \thanks{This material is based upon work supported by the National Science Foundation Graduate Research Fellowship Program under Grant Number DGE-1144245.}}
\address{University of Illinois at Urbana-Champaign}

\begin{document}

\ninept

\title{Delay-Performance Tradeoffs in Causal Microphone Array Processing}
\maketitle
\begin{abstract}
In real-time listening enhancement applications, such as hearing aid signal processing,
sounds must be processed with no more than a few milliseconds of
delay to sound natural to the listener. Listening devices can achieve
better performance with lower delay by using microphone arrays to
filter acoustic signals in both space and time. Here, we analyze the
tradeoff between delay and squared-error performance of causal multichannel
Wiener filters for microphone array noise reduction. We compute exact
expressions for the delay-error curves in two special cases and present
experimental results from real-world microphone array recordings.
We find that delay-performance characteristics are determined by both
the spatial and temporal correlation structures of the signals.
\end{abstract}

\section{Introduction}

Listening enhancement applications, such as hearing aid processing \cite{kates2008aids}
and audio augmented reality \cite{valimaki2015assisted}, differ from
other audio enhancement applications, like teleconferencing and speech
recognition, in part because of their strict delay constraints. Since
users hear both live and processed signals simultaneously, these systems must process sound with no more than a few milliseconds
of delay. Discerning listeners can notice delays as low as 3 ms and
are disturbed by delays greater than 10 ms \cite{agnew2000just}.
Listeners with hearing loss can tolerate greater delay, around 20
ms for closed-fitting hearing aids \cite{stone1999tolerable} and
6 ms for open-fitting hearing aids \cite{stone2008tolerable}. Delays
longer than about 30 ms can impair the user's ability to speak \cite{stone2002tolerable}. 

This delay requirement limits the performance of audio enhancement systems.
In single-channel systems, the frequency resolution of a frequency-selective filter generally
improves with longer delay. Modern single-microphone audio enhancement algorithms \cite{makino2018audio}, such as those employing
time-frequency masks \cite{yilmaz2004duet} and non-negative matrix
factorization \cite{ozerov2010multichannel}, often process speech
using short-time Fourier transform (STFT) frames of 60 ms or longer to
maximize time-frequency sparsity \cite{yilmaz2004duet}. These algorithms
are effective in many applications, but their delay is too large for
listening enhancement. 

Multichannel audio enhancement systems use microphone arrays to spatially
separate signals \cite{benesty2008microphone,brandstein2013microphone,gannot2017consolidated}.
Many multichannel methods are also applied in the STFT domain to more
easily model reverberation \cite{gannot2017consolidated,pedersen2008convolutive}.
In principle, however, spatial processing should require
minimal delay: for example, a linear array can enhance a source at
broadside with zero delay by simply summing its inputs. Whereas the
frequency resolution of a temporal filter depends on its duration, the spatial
resolution of an array is determined by its spatial extent. Multichannel
listening systems can use both spatial and spectral diversity to separate
signals. It is natural to ask, therefore, whether devices with large
arrays can enhance audio with lower delay than those with small arrays.
That is, \emph{can we use array processing to trade space for time?}

There is a large body of literature on array processing for listening
devices, e.g. \cite{doclo2015magazine,doclo2008acoustic}, and causal
multichannel filters have been studied in the contexts of dereverberation
\cite{naylor2010speech,nakatani2008blind,schwartz2015online,benesty2007microphone}
and noise and echo control \cite{hansler2005acoustic}. In \cite{chen2008minimum},
the authors considered the minimum filter delay required to cover
the full aperture of an array. There have also been several proposed low-delay single-microphone
filtering and source separation techniques \cite{kates2005multichannel,lollmann2009low,barker2015low}. However, to the best of our knowledge,
there has been no prior study of delay-performance tradeoffs in array
processing. 

\begin{figure}
\begin{centering}
\includegraphics{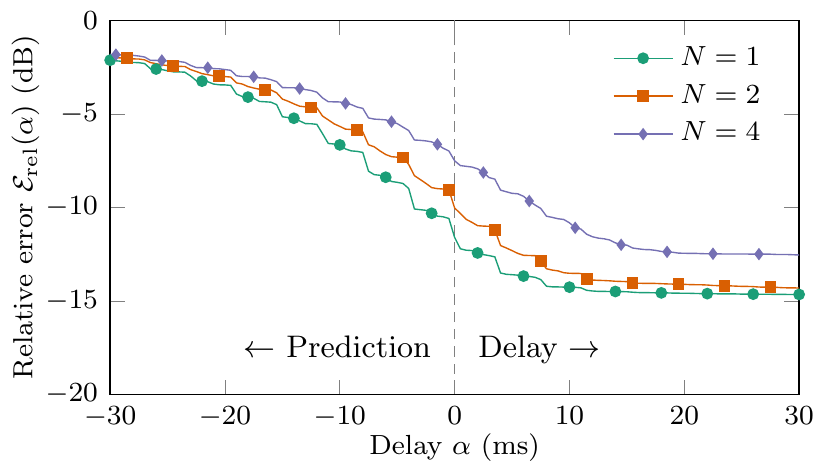}
\par\end{centering}
\caption{\label{fig:vowels}Relative MSE as a function of delay for isolating one source from a mixture of $N$ synthetic
speechlike sounds (see Section 3) and uncorrelated noise using single-channel Wiener filters.}
\end{figure}

Here we approach audio enhancement as a stationary linear estimation
problem: given an observed signal from the infinite past to time $t$,
what is the linear minimum mean square error (MSE) estimate of a desired
signal at time $t-\alpha$? Positive values of $\alpha$ correspond
to delay and negative values to prediction. Such problems are well
understood in the scalar case: for certain signals, we can use spectral
factorization to compute exact expressions for the MSE as a function
of $\alpha$ \cite{wiener1949extrapolation,bode1950simplified,van2004detection1}.
For example, Figure \ref{fig:vowels} shows delay-error curves
for separating several spectrally distinct speechlike sounds, which
will be described in Section \ref{fig:Cocktail-party-experiment}.
As $\alpha$ increases, the MSE decreases from the variance of the
target signal to the MSE of a noncausal Wiener filter. We can apply
similar theoretical tools in the multivariate case \cite{wiener1958prediction,wong1961multidimensional}
to analyze delay-performance tradeoffs for causal multichannel Wiener
filters (CMWF) in terms of the spatial and temporal correlation structures
of the source signals. In this work, we will derive a general expression
for the MSE performance of a CMWF as a function of $\alpha$, find
exact expressions for idealized mixing models, and present experimental
results from wearable and distributed microphone arrays in a real
room.

\section{Delay-Constrained Multichannel Filtering}

Consider a mixture of $N$ sources captured by $M$ microphones. Let
the sources $\boldsymbol{s}(t)=\left[s_{1}(t),\dots,s_{N}(t)\right]^{T}$
and additive noise $\boldsymbol{z}(t)=\left[z_{1}(t),\dots,z_{M}(t)\right]^{T}$
be wide-sense stationary continuous-time random processes that are
uncorrelated with each other. Let $a_{m,n}(t)$, $m=1,\dots, M$, $n=1,\dots,N$
be known causal impulse responses and let $\boldsymbol{w}_{\alpha}^{T}(t)=\left[w_{\alpha,1}(t),\dots,w_{\alpha,M}(t)\right]$
be filter impulse responses. Denote the observed signals by $\boldsymbol{x}(t)=\left[x_{1}(t),\dots,x_{M}(t)\right]^{T}$
and the system output by $y_{\alpha}(t)$, where 
\begin{align}
x_{m}(t) & =\sum_{n=1}^{N}(a_{m,n}*s_{n})(t)+z_{m}(t),\quad m=1,\dots,M,\text{ and}\\
y_\alpha(t) & =\sum_{m=1}^{M}\left(w_{\alpha,m}*x_{m}\right)(t),
\end{align}
and $*$ denotes linear convolution. We define the desired output
signal $d_{\alpha}(t)$ to be the first source as captured by the
first microphone\textemdash for example, a target talker reproduced
at the microphone nearest the listener's ear\textemdash and delayed
by time $\alpha$:
\begin{equation}
d_{\alpha}(t)=\left(a_{11}*s_{1}\right)(t-\alpha).
\end{equation}

To understand fundamental tradeoffs in performance, we restrict our
attention to the best-case scenario in which all signals are stationary
in both space and time and have known statistics. Let $\boldsymbol{A}(\omega)$
be the $M\times N$ frequency response matrix corresponding to the
$a_{m,n}(t)$'s. Let $\boldsymbol{r}_{s}(t)$, $\boldsymbol{r}_{z}(t)$,
$r_{d}(t)$, and $\boldsymbol{r}_{x}(t)$ be the autocorrelation sequences
of the corresponding random variables and let $\boldsymbol{R}_{s}(\omega)$,
$\boldsymbol{R}_{z}(\omega)$, $R_{d}(\omega)=|A_{1,1}(\omega)|^{2}R_{s_{1}}(\omega)$,
and $\boldsymbol{R}_{x}(\omega)=\boldsymbol{A}(\omega)\boldsymbol{R}_{s}(\omega)\boldsymbol{A}^{H}(\omega)+\boldsymbol{R}_{z}(\omega)$
be their respective Fourier transforms. To ensure that the CMWF is
well defined, we assume that $\boldsymbol{R}_{x}(\omega)$ is positive
definite for all $\omega$ of interest. Let $\boldsymbol{r}_{xd}(t)$
be the cross-correlation of $\boldsymbol{x}(t)$ with $d_{0}(t)$
and let $\boldsymbol{R}_{xd}(\omega)=\boldsymbol{A}_{1}(\omega)R_{s_{1}}(\omega)A_{1,1}^{*}(\omega)$
be its Fourier transform, where $\boldsymbol{A}_{1}(\omega)$ is the
column of $\boldsymbol{A}(\omega)$ corresponding to the target source.
Let $\boldsymbol{W}_{\alpha}^{T}(\omega)$ be the Fourier transform
of $\boldsymbol{w}_{\alpha}^{T}(t)$.

\subsection{Causal filter performance}

The CMWF $\boldsymbol{w}_{\alpha}^{T}(t)$ must satisfy the
Wiener-Hopf equation \cite{wiener1949extrapolation},
\begin{equation}
\boldsymbol{r}_{xd}^{T}(t-\alpha)=\int_{0}^{\infty}\boldsymbol{w}_{\alpha}^{T}(u)\boldsymbol{r}_{x}(t-u)\,\mathrm{d}u,\quad 0<t<\infty.\label{eq:wiener-hopf}
\end{equation}
The MSE between $y_\alpha(t)$ and $d_\alpha(t)$ is 
\begin{equation}
\mathcal{E}(\alpha)=r_{d}(0)-\int_{-\infty}^{\infty}\boldsymbol{w}_{\alpha}^{T}(t)\boldsymbol{r}_{xd}(t-\alpha)\,\mathrm{d}t.\label{eq:wiener_mse}
\end{equation}
The \emph{noncausal} ($\alpha \longrightarrow \infty$) solution to (\ref{eq:wiener-hopf}) and its error power
are readily expressed in the frequency domain:
\begin{align}
\boldsymbol{W}_{\mathrm{nc}}^{T}(\omega) & =\boldsymbol{R}_{xd}^{H}(\omega)\boldsymbol{R}_{x}^{-1}(\omega)\\
\mathcal{E}_{\mathrm{nc}} & =\int_{-\infty}^{\infty}\left[R_{d}(\omega)-\boldsymbol{R}_{xd}^H(\omega)\boldsymbol{R}_{x}^{-1}(\omega)\boldsymbol{R}_{xd}(\omega)\right]\frac{\mathrm{d}\omega}{2\pi}.
\end{align}

For finite $\alpha$, we
can solve (\ref{eq:wiener-hopf}) by first decomposing $\boldsymbol{R}_{x}(\omega)$
into its \emph{spectral factors} \cite{wiener1958prediction},
\begin{equation}
\boldsymbol{R}_{x}(\omega)=\boldsymbol{G}(\omega)\boldsymbol{G}^{H}(\omega),\label{eq:factorization}
\end{equation}
where $\boldsymbol{G}(\omega)$ and its inverse are both causal. We
proceed by decorrelating $\boldsymbol{x}(t)$ using $\boldsymbol{G}^{-1}(\omega)$
and then solving (\ref{eq:wiener-hopf}) for the decorrelated signals
\cite{wong1961multidimensional} to find the causal filter 
\begin{equation}
\boldsymbol{W}_{\alpha}^{T}(\omega)=\left[e^{-j\omega\alpha}\boldsymbol{R}_{xd}^{H}(\omega)(\boldsymbol{G}^{H}(\omega))^{-1}\right]_{+}\boldsymbol{G}^{-1}(\omega),\label{eq:nc_filter}
\end{equation}
where $\left[\cdot\right]_{+}$ denotes the causal part of the argument,
that is, time-domain truncation from $t=0$. Let $\tilde{\boldsymbol{R}}^{T}(\omega)=\boldsymbol{R}_{xd}^{H}(\omega)(\boldsymbol{G}^{H}(\omega))^{-1}$.
For the listening enhancement application, this vector can be written
\begin{equation}
\tilde{\boldsymbol{R}}^{T}(\omega)=A_{1,1}(\omega)R_{s_{1}}(\omega)\boldsymbol{A}_{1}^{H}(\omega)(\boldsymbol{G}^{H}(\omega))^{-1}.\label{eq:correlation}
\end{equation}

Let $\tilde{\boldsymbol{r}}^{T}(t)$ be the inverse Fourier transform
of $\tilde{\boldsymbol{R}}^{T}(\omega)$. Substituting $\boldsymbol{w}_{\alpha}^{T}$
from (\ref{eq:nc_filter}) into (\ref{eq:wiener_mse}), using the
spectral factorization (\ref{eq:factorization}) and Parseval's identity,
and rearranging terms \cite{van2004detection1}, we can show that
\begin{equation}
\boxed{\mathcal{E}(\alpha)=\mathcal{E}_{\mathrm{nc}}+\int_{-\infty}^{-\alpha}\tilde{\boldsymbol{r}}^{T}(t)\tilde{\boldsymbol{r}}(t)\,\mathrm{d}t.}\label{eq:causal_error}
\end{equation}
Thus, \emph{the error penalty due to causality is the energy in $\tilde{\boldsymbol{r}}(t)$
for $t<-\alpha$. }Our goal is to understand how $\mathcal{E}(\alpha)$
depends on the spatial and spectral characteristics of the source signals.
While multivariate spectral factorizations are often difficult to
compute in practice \cite{kucera1991factorization}, we can find exact
expressions for certain special cases that provide insight about the
delay-constrained array processing problem.

\subsection{Uniform linear array}

\medmuskip=1mu
\arraycolsep=3pt

First, consider a plane wave incident upon a uniform linear array
of $M$ sensors with the reference at one end. Let $\tau$ be the
time difference of arrival (TDOA) between adjacent microphones, let
$R_{s}(\omega)=1$ and let $\boldsymbol{R}_{z}(\omega)=\sigma^{2}\boldsymbol{I}$,
so that
\begin{align}
\boldsymbol{R}_{xd}^{H} & =\left[\begin{matrix}1 & e^{+j\omega\tau} & \cdots & e^{+j\omega(M-1)\tau}\end{matrix}\right]\,\text{ and }\\
\boldsymbol{R}_{x}(\omega) & \!=\!\left[\begin{matrix}\sigma^{2}+1 & e^{+j\omega\tau} & \!\cdots\! & e^{+j\omega(M-1)\tau}\\
e^{-j\omega\tau} & \sigma^{2}+1 &  & e^{+j\omega(M-2)\tau}\\
\vdots &  & \!\ddots\! & \vdots\\
e^{-j\omega(M-1)\tau} & e^{-j\omega(M-2)\tau} & \!\cdots\! & \sigma^{2}+1
\end{matrix}\right].
\end{align}

A convenient spectral factor is the lower triangular matrix 
\begin{equation}
\boldsymbol{G}(\omega)\!=\!\left[\begin{matrix}b_{1}(\sigma^{2}+1) & 0 & \cdots & 0\\
b_{1}e^{-j\omega\tau} & b_{2}(\sigma^{2}+2) &  & 0\\
\vdots &  & \!\ddots\! & \vdots\\
b_{1}e^{\!-j\omega(M\!-\!1)\tau} & b_{2}e^{\!-j\omega(M\!-\!2)\tau} & \!\cdots\! & b_{M}(\sigma^{2}+M)
\end{matrix}\right]
\end{equation}
where $b_{m}=\sqrt{\sigma^{2}/((\sigma^{2}+m)(\sigma^{2}+m-1))}$.
Applying (\ref{eq:correlation}) and taking the inverse Fourier transform, we have
\begin{align}
\tilde{\boldsymbol{r}}^{T}(t)=\left[\begin{matrix}b_{1} & b_{2}\delta(t+\tau) & \cdots & b_{M}\delta(t+(M-1)\tau)\end{matrix}\right].
\end{align}
Finally, from (\ref{eq:causal_error}), the MSE is
\begin{align}
\mathcal{E}(\alpha) & =\frac{\sigma^{2}}{\sigma^{2}+M}+\sum_{m=0}^{M-1}b_{m+1}^{2}u\left(m\tau-\alpha\right)\\
 & =\frac{\sigma^{2}}{\sigma^{2}+\sum_{m=0}^{M-1}\bar{u}(\alpha-m\tau)},
\end{align}
where $u(t)=1$ if $t>0$ and $\bar{u}(t)=1$ if $t\ge0$. Thus, the
error is reduced for each microphone that the source reaches within
time $\alpha$ of reaching the reference. The delay-error curve is
a piecewise constant function with steps of width $|\tau|$ and decreasing
heights that depend on $\sigma^{2}$. 

\subsection{Two-source, two-microphone separation}

\begin{figure}
\begin{centering}
\includegraphics{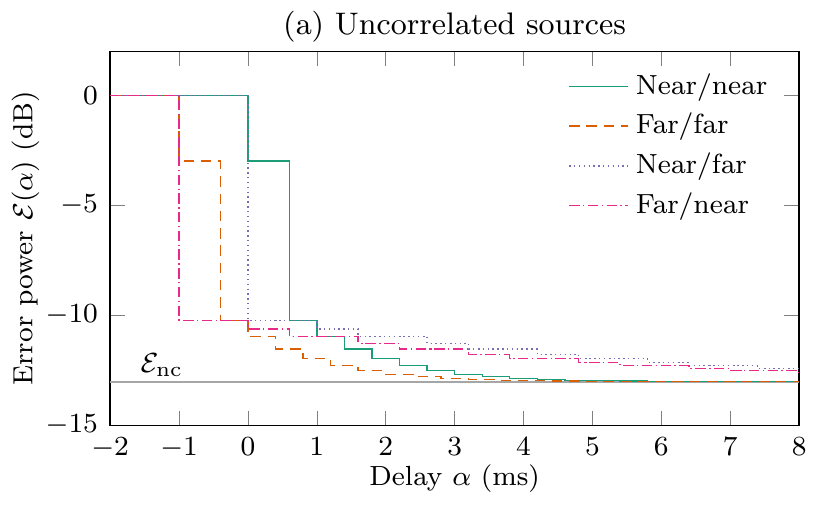}
\par\end{centering}
\begin{centering}
\includegraphics{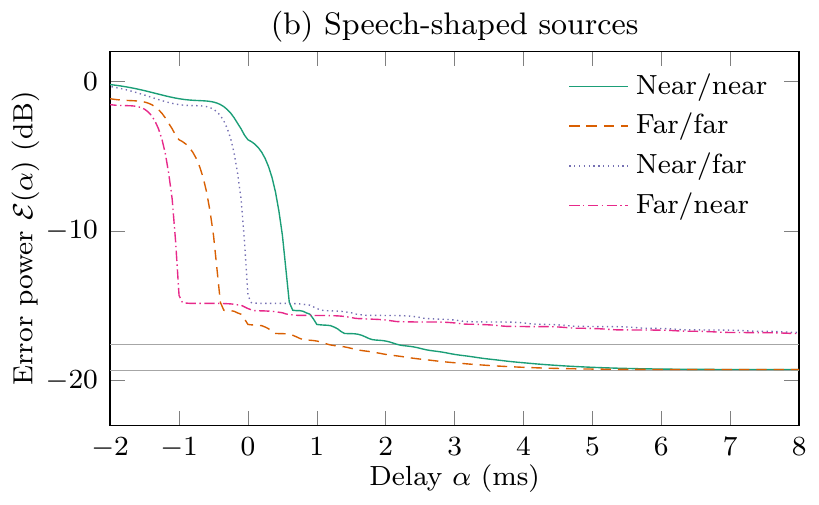}
\par\end{centering}
\caption{\label{fig:anechoic}Delay-error curves for two plane wave sources
and two sensors with $|\tau_{1}|=1$ ms, $|\tau_{2}|=$0.6 ms, and
$\sigma^{2}=-20$ dB. The legend indicates the placement of the target/interference
sources with respect to the reference.}
\end{figure}

We can follow a similar procedure with multiple sources. Consider a scenario
with two plane wave sources and two microphones. Let $\tau_{1}$ and
$\tau_{2}\ne\tau_{1}$ be the TDOAs of the sources, let \textbf{$\boldsymbol{R}_{s}(\omega)=\boldsymbol{I}$}
and let $\boldsymbol{R}_{z}(\omega)=\sigma^{2}\boldsymbol{I}$ with
$\sigma^{2}>0$, so that 

\begin{align}
\boldsymbol{R}_{xd}^{H}(\omega) & =\left[\begin{matrix}1 & e^{+j\omega\tau_{1}}\end{matrix}\right],\:\text{ and}\\
\boldsymbol{R}_{x}(\omega) & =\left[\begin{matrix}2+\sigma^{2} & e^{+j\omega\tau_{1}}+e^{+j\omega\tau_{2}}\\
e^{-j\omega\tau_{1}}+e^{-j\omega\tau_{2}} & 2+\sigma^{2}
\end{matrix}\right].
\end{align}

The determinant of $\boldsymbol{R}_{x}(\omega)$ can be written
\begin{equation}
\mathrm{det}\boldsymbol{R}_{x}(\omega)=\gamma^{-1}\left|1-\gamma e^{-j\omega(\tau_{1}-\tau_{2})}\right|^{2},
\end{equation}
where $\gamma$ is a scalar that depends only on $\sigma^{2}$. The
spectral factorization of $\boldsymbol{R}_{x}(\omega)$ takes different
forms depending on the signs of $\tau_{1}$ and $\tau_{2}$, but $\tilde{\boldsymbol{R}}^{T}(\omega)$
always includes a term of the form $(1-\gamma e^{+j\omega|\tau_{1}-\tau_{2}|})^{-1}$,
which results in an infinite-duration $\tilde{\boldsymbol{r}}^{T}(t)$.
Applying (\ref{eq:causal_error}), we find that 
\begin{equation}
\mathcal{E}(\alpha)=\begin{cases}
\mathcal{E}_{\mathrm{nc}}+\frac{u(t_{0}\!-\!\alpha)+c_{1}^{2}\gamma u(t_{1}\!-\!|\tau_{1}\!-\!\tau_{2}|-\alpha)+c_{2}^{2}f(t_{1})}{\sigma^{2}+2}, & \text{if }\tau_{1}\tau_{2}>0\\
\mathcal{E}_{\mathrm{nc}}+\sqrt{\gamma}u(t_{0}\!-\!\alpha)+f(t_{0}\!-\!|\tau_{1}|)+\gamma f(t_{1}), & \text{if }\tau_{1}\tau_{2}\le0
\end{cases}
\end{equation}
where $t_{0}=\min(0,\tau_{1})$, $t_{1}=\max(0,\tau_{1},\tau_{2},\tau_{1}-\tau_{2})$,
\begin{equation}
f(t) =\gamma^{1+2\max(0,\lfloor(\alpha-t)/|\tau_{2}-\tau_{1}|\rfloor+1)}/(1-\gamma^{2}),\,\text{ and}\\
\end{equation}
\begin{equation}
(c_{1},c_{2}) =\begin{cases}
(0,0), & \text{if }|\tau_{1}|=|\tau_{2}|\\
\left(\sigma^{2}+1,\gamma+\gamma\sigma^{2}-1\right), & \text{if }|\tau_{1}|<|\tau_{2}|\\
\left(1,\sigma^{2}+1-\gamma\right), & \text{if }|\tau_{1}|>|\tau_{2}|.
\end{cases}
\end{equation}
This delay-error curve is also piecewise constant, but has a geometric
``tail'' that decays with a rate of roughly $\gamma^{2/|\tau_{2}-\tau_{1}|}$.
The height of the steps is determined by $\sigma^{2}$ and the width
is determined by $\left|\tau_{2}-\tau_{1}\right|$, which depends
on the distance between the sources. For large positive $\alpha$,
$\mathcal{E}(\alpha)$ approaches $\mathcal{E}_{\mathrm{nc}}$.

Figure \ref{fig:anechoic}(a) shows $\mathcal{E}(\alpha)$ for four
combinations of source placement. The causality penalty takes a different
form depending on the relative placement of sources and microphones.
For example, if both the target and interference source are closer
to microphone 1 than microphone 2 (near/near), then the second microphone
does not contribute any information at $\alpha=0$. If the sources
are on opposite sides, then the difference in TDOAs, $|\tau_{1}-\tau_{2}|$
is larger, and therefore $\mathcal{E}(\alpha)$ decays more slowly.

\subsection{Temporally correlated signals}

The expressions above were derived for uncorrelated source and noise
processes. In many applications, however, the signals of interest
are correlated and can therefore be separated spectrally as well as
spatially. It is difficult in general to predict the effects of signal
correlation on the delay-error curve. However, if the entries of $\boldsymbol{R}_{x}(\omega)$
share a common spectral factor\textemdash for example, if the sources
are identically distributed and are recorded by identical microphones\textemdash then
we can write $\boldsymbol{R}_{x}(\omega)=H(\omega)\hat{\boldsymbol{G}}(\omega)\hat{\boldsymbol{G}}^{H}(\omega)H^{*}(\omega)$
and $\boldsymbol{R}_{xd}^{H}(\omega)=H(\omega)\hat{\boldsymbol{R}}_{xd}^{H}(\omega)H^{*}(\omega)$,
where $H(\omega)H^{*}(\omega)$ is the scalar spectral factorization
of the common factor. Then we have
\begin{align}
\boldsymbol{R}_{xd}^{H}(\omega)(\boldsymbol{G}^{H}(\omega))^{-1} & =H(\omega)\hat{\boldsymbol{R}}_{xd}^{H}(\omega)(\hat{\boldsymbol{G}}^{H}(\omega))^{-1}\\
\tilde{\boldsymbol{r}}^{T}(t) & =(h*\hat{\tilde{\boldsymbol{r}}}^{T})(t).
\end{align}

Since $h(t)$ is causal, it spreads the energy of $\tilde{\boldsymbol{r}}(t)$
forward in time. Figure \ref{fig:anechoic}(b) shows the same scenario
as in the previous section, but with identically distributed speech-shaped
sources. The error is lower overall, the steps are smoother, and the
filter can begin to separate the signals even before they reach either
microphone. 

\section{Experiments\label{sec:Moving-Wearable-Arrays}}

\begin{figure}
\begin{centering}
\hfill{}\includegraphics{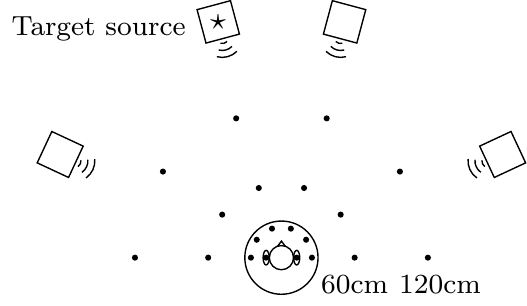}\hfill{}\includegraphics[height=3.2cm]{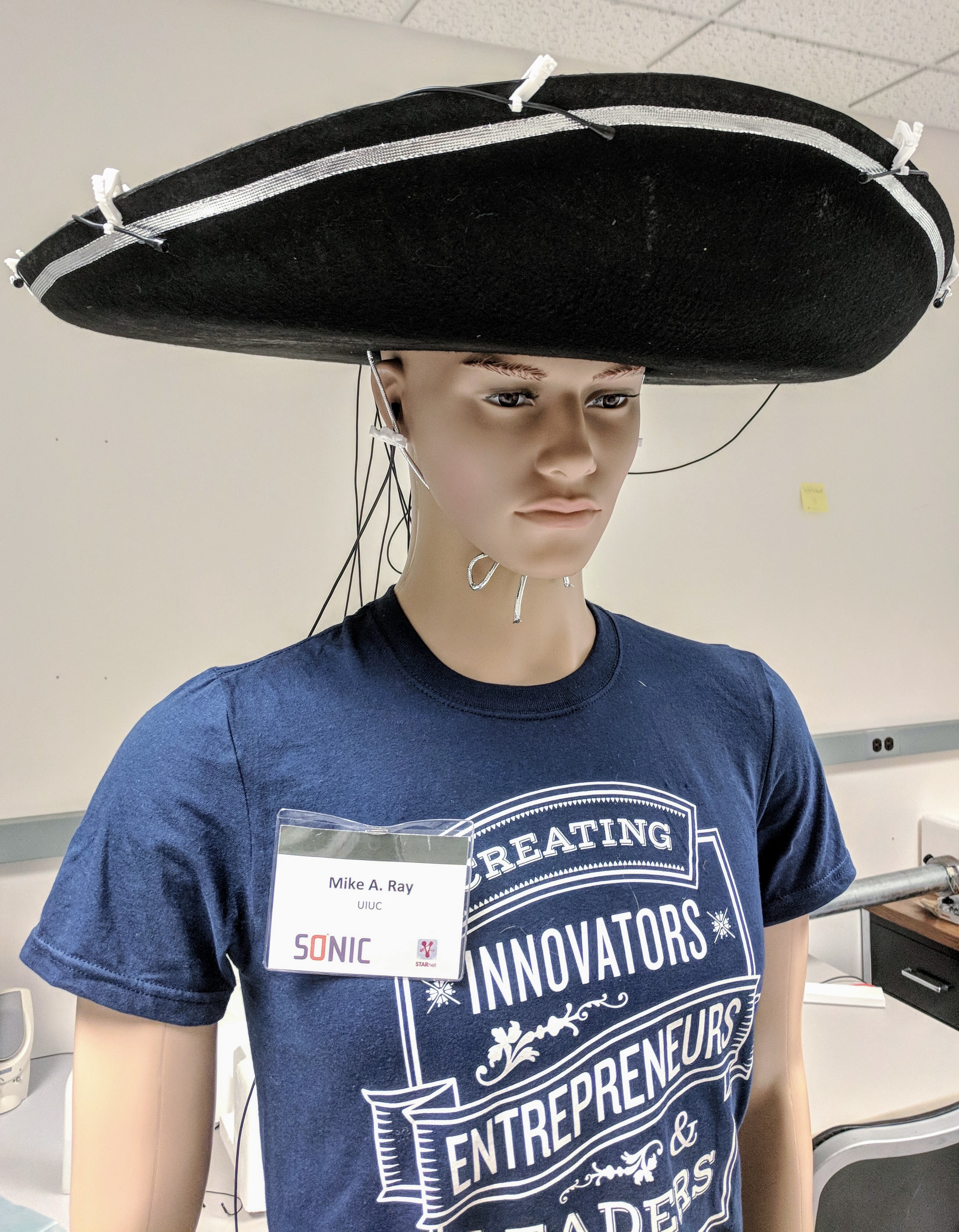}\hfill{}
\par\end{centering}
\caption{\label{fig:Cocktail-party-experiment}Left: Recording setup. Circles
are microphones and squares are loudspeakers. Right: Hat-mounted microphone
array.}
\end{figure}

To evaluate delay-performance tradeoffs in realistic conditions, we
recorded audio mixtures using a wearable microphone array in a cocktail
party scenario at the Augmented Listening Laboratory at the University
of Illinois at Urbana-Champaign, which has a reverberation time of
around $T_{60}=300$ ms. The recording setup, shown in Figure
\ref{fig:Cocktail-party-experiment}, consisted of twenty omnidirectional
lavalier microphones: two at the left and right ears of a mannequin
``listener,'' six along the perimiter of a hat with radius 30 cm,
and twelve mounted on stands at 60 cm and 120 cm distances from
the listener. The reference microphone is that in the left ear. Source
signals were produced by loudspeakers two meters away from the listener.
The acoustic impulse responses between the loudspeakers and microphones
were measured using linear sweeps. All data was sampled at 16 kHz.

The signals were separated using the discrete-time, finite-length
version of the CMWF. Let $\bar{\boldsymbol{x}}[k]=\left[\boldsymbol{x}^{T}[k],\dots,\boldsymbol{x}^{T}[k-L+1]\right]^{T}$
and $\boldsymbol{\bar{w}}_{\alpha}^{T}=\left[\boldsymbol{w}_{\alpha}^{T}[0],\dots,\boldsymbol{w}_{\alpha}^{T}[L-1]\right]$
be stacked vectors of the sampled multichannel signals and the finite
impulse response filter coefficients, respectively. Let $y_\alpha[k]=\bar{\boldsymbol{w}}_{\alpha}^{T}\bar{\boldsymbol{x}}[k]$
be the filter output sequence and let $d_{\alpha}[k]$ be the desired
output sequence. Let $\bar{\boldsymbol{r}}_{x}=\mathbb{E}\left[\bar{\boldsymbol{x}}[n]\bar{\boldsymbol{x}}^{T}[n]\right]$
and $\bar{\boldsymbol{r}}_{xd}(\alpha)=\mathbb{E}\left[\bar{\boldsymbol{x}}[n]d_{\alpha}[n]\right]$,
where $\mathbb{E}[\cdot]$ is expectation. The linear minimum MSE filter
coefficients are \cite{benesty2008microphone}
\begin{align}
\bar{\boldsymbol{w}}_{\alpha}^{T} & =\boldsymbol{\bar{r}}_{xd}^{T}(\alpha)\bar{\boldsymbol{r}}_{x}^{-1}.
\end{align}
In our experiments, $\bar{\boldsymbol{r}}_{x}$ was computed using
truncated impulse response measurements. We applied diagonal loading comparable
to the source power to account for modeling errors and ambient noise. We used discrete-time filters with length $L=2048$
samples (128 ms). For each experiment we report the sample MSE relative
to the source power, computed as $\mathcal{E}_{\mathrm{rel}}(\alpha)=10\log_{10}\sum_{k}\left(y_\alpha[k]-d_{\alpha}[k]\right)^{2}/\sum_{k}d_{\alpha}^{2}[k]$.

\begin{figure}
\begin{centering}
\includegraphics{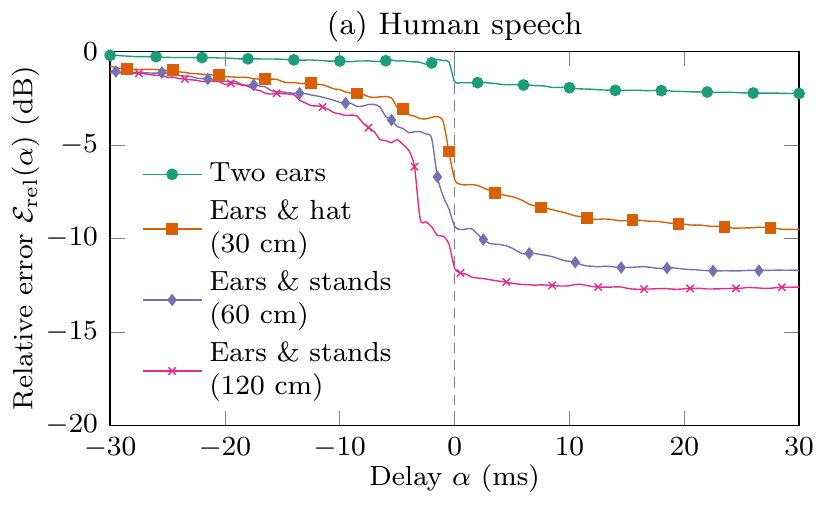}
\par\end{centering}
\begin{centering}
\includegraphics{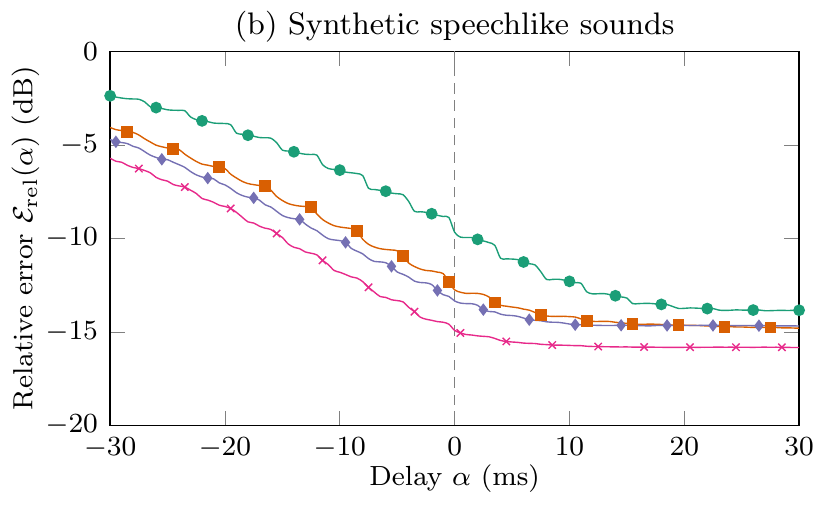}
\par\end{centering}
\caption{\label{fig:cpp_results}Experimental delay-error results for isolating a single target source from a mixture of four sources.}
\end{figure}

Figure \ref{fig:cpp_results}(a) shows delay-error curves for four
simultaneous talkers using arrays of up to eight microphones at varying
distances. The speech signals were twenty-second clips taken from
the VCTK dataset \cite{veaux2017cstr} and the filters were designed
using a single approximate long-term average speech autocorrelation. Because we model the signals as identically distributed,
the filters must rely on spatial rather than spectral diversity to
separate the sources. As the radius of the array increases, the curves
move downward and to the left, indicating that the larger-aperture
arrays can achieve similar performance with lower delay compared to
the smaller-aperture arrays. In fact, since the source signals reach
the microphone stands several milliseconds before they reach the listener,
the system could operate with negative delay. 

The two-channel filter performs poorly in this experiment because
it does not take advantage of the time-frequency sparsity of speech
signals, which many speech enhancement algorithms exploit. To account
for the benefits of sparsity within the stationary estimation framework
of this paper, we repeated the experiment with four stationary speechlike
sounds generated using the Vocaloid music synthesis software. Each
ten-second source signal represents a different vowel sung in a different
key. Although the signals are deterministic and periodic, the filters
were designed based on 50 ms von Hann-windowed autocorrelation sequences.
Figure \ref{fig:vowels} shows the delay-error curves for single-channel
mixtures of these sources and Figure \ref{fig:cpp_results}(b) compares
the separation performance of multichannel filters with different
array sizes. Because the sources are approximately disjoint in the
frequency domain, a one- or two-channel filter can separate them effectively,
but requires a delay to do so. The larger microphone arrays also benefit
from longer delay, but perform better for small $\alpha$. For example, the performance
of the hat-mounted array with zero delay matches that of the binaural
microphones with about 10 ms delay, which would be perceptible to
many listeners.

\section{Conclusions}

The theoretical and experimental results presented here suggest that
larger arrays can separate sound sources with lower delay and that
the delay-performance tradeoff depends on both the spatial and temporal
correlation structure of the observed signals. When microphones are
located between the listener and sound sources, those sensors receive
the signals before the listening device, shifting the delay-performance
curve to the left. Arrays also provide spatial gain, which improves
overall performance regardless of delay. When signals are spectrally
distinct, a single-channel filter could separate them effectively
given a long enough delay, but an array can achieve the same performance
with little or no delay.

Much remains to be understood about delay-constrained array processing.
For example, equations (\ref{eq:correlation}) and (\ref{eq:causal_error})
tell us little in general about the effects of reverberation and signal
spectra on delay. Furthermore, because many signals of interest are
nonstationary, we must also consider time-varying causal array processing.
Finally, to realize the benefits of spatial diversity in delay-constrained
listening enhancement, listening devices must use larger microphone
arrays than they do today. Large wearable and distributed arrays could
allow us to apply stronger noise reduction while meeting the strict
delay constraints of real-time listening applications.

\bibliographystyle{ieeetr}
\bibliography{references}

\end{document}